\documentclass[10pt,conference]{IEEEtran}
\IEEEoverridecommandlockouts
\usepackage{cite}
\usepackage{amsmath,amssymb,amsfonts}
\usepackage{algorithmic}
\usepackage{graphicx}
\usepackage{textcomp}
\usepackage{xcolor}
\usepackage{hyperref}
\def\BibTeX{{\rm B\kern-.05em{\sc i\kern-.025em b}\kern-.08em
    T\kern-.1667em\lower.7ex\hbox{E}\kern-.125emX}}
\begin{document}

\title{\textit{COMEX}: A Tool for Generating Customized Source Code Representations}


\author{%
  \IEEEauthorblockN{%
    Debeshee Das\IEEEauthorrefmark{1}\textsuperscript{\textsection},
    Noble Saji Mathews\IEEEauthorrefmark{1}\textsuperscript{\textsection},
    Alex Mathai\IEEEauthorrefmark{2},\\
    Srikanth Tamilselvam\IEEEauthorrefmark{2},
    Kranthi Sedamaki\IEEEauthorrefmark{1},
    Sridhar Chimalakonda\IEEEauthorrefmark{1} and
    Atul Kumar\IEEEauthorrefmark{2}%
  }%
  \IEEEauthorblockA{\IEEEauthorrefmark{1} 
Indian Institute of Technology Tirupati, India
  }%
  \IEEEauthorblockA{\IEEEauthorrefmark{2} IBM Research, India}%
  \IEEEauthorblockA{
\{debesheedas, elbonleon, alexmathai98, srikanthtamilselvam, skranthi4444, sridhar.chimalakonda, atulkumar\}@gmail.com
}
}

\maketitle
\begingroup\renewcommand\thefootnote{\textsection}
\footnotetext[1]{Authors have contributed equally}
\endgroup

\begin{abstract}
Learning effective representations of source code is critical for any Machine Learning for Software Engineering (ML4SE) system. Inspired by natural language processing, large language models (LLMs) like \textit{Codex} and \textit{CodeGen} treat code as generic sequences of text and are trained on huge corpora of code data, achieving state of the art performance on several software engineering (SE) tasks. However, valid source code, unlike natural language, follows a strict structure and pattern governed by the underlying grammar of the programming language. Current LLMs do not exploit this property of the source code as they treat code like a sequence of tokens and overlook key structural and semantic properties of code that can be extracted from code-views like the Control Flow Graph (CFG), Data Flow Graph (DFG), Abstract Syntax Tree (AST), etc. Unfortunately, the process of generating and integrating code-views for every programming language is cumbersome and time consuming. To overcome this barrier, we propose our tool \texttt{COMEX} - a framework that allows researchers and developers to create and combine multiple code-views which can be used by machine learning (ML) models for various SE tasks. Some salient features of our tool are: (i) it works directly on source code (which need not be compilable), (ii) it currently supports Java and C\#, (iii) it can analyze both method-level snippets and program-level snippets by using both intra-procedural and inter-procedural analysis, and (iv) it is easily extendable to other languages as it is built on \emph{tree-sitter} - a widely used incremental parser that supports over 40 languages. We believe this easy-to-use code-view generation and customization tool will give impetus to research in source code representation learning methods and ML4SE. The demonstration of our tool can be found at \url{https://youtu.be/GER6U87FVbU}.

\end{abstract}

\begin{IEEEkeywords}
Representation Learning, Static Analysis
\end{IEEEkeywords}

\section{Introduction}
Source code representation learning is the task of effectively capturing useful syntactic and semantic information
embedded in source code \cite{allamanis2017learning}. It forms the backbone of ML pipelines for various SE tasks such as \textit{code classification, bug prediction, code clone detection} and \textit{code summarization}. Therefore, representing source code for use in ML models, with minimal loss of important information is an active research area \cite{10.1145/3212695}. It is important to note that source code is different from natural language as it follows an unambiguous structure and pattern, usually adhering to a strict underlying grammar. Hence, while creating representations for source code, it is important to infuse information from this unique structural aspect. To address this, many works including GraphCodeBERT\cite{guo2020graphcodebert} and GREAT\cite{49316} have explored leveraging code-views as a means to learn source code representations. Unfortunately, the process of generating code-views for multiple programming languages and customizing them for various SE tasks is often a time consuming process. 

Most available tools are (a) positioned for analysis on compiled or compilable code (and not incomplete or uncompilable source code), (b) are specific for a single language, and (c) are not able to support both intra-procedural and inter-procedural analysis. 
To address these concerns, we propose \texttt{COMEX} - a framework that (a) works directly on source code to generate and combine multiple code-views, (b) supports Java and C\# (with planned support for other languages) and (c) works for both method-level and program-level snippets using intra-procedural and inter-procedural analysis. Since it is based on a single parser package (\textit{tree-sitter}\footnote{\url{https://tree-sitter.github.io/tree-sitter/}}), it can be extended to new languages without additional dependencies. 

As of today, most state-of-the-art models like \textit{CodeGen} \cite{nijkamp2023codegen} and \textit{Codex} \cite{chen2021evaluating} treat source code like free flowing text. Though this assumption helps simplify the required data pre-processing, it loses out on many structural aspects of code. Recently, works like \textit{NSG} \cite{mukherjee2021neural} have shown the benefits of using code structure. \textit{NSG} leverages weak supervision using a syntax tree to generate full-length syntactically valid method bodies. Their results showcase that using this technique, even a small model ($63$ million parameters) can outperform LLMs like \textit{Codex} ($12$ billion parameters). 
To fuel research on similar grounds, we hope that with this package, we have lowered the entry barrier for researchers to easily integrate and leverage code-views while learning source code representations.


\section{Related Work}
    \begin{figure*}[h]
    \centering
    {\includegraphics[width=6.5in]{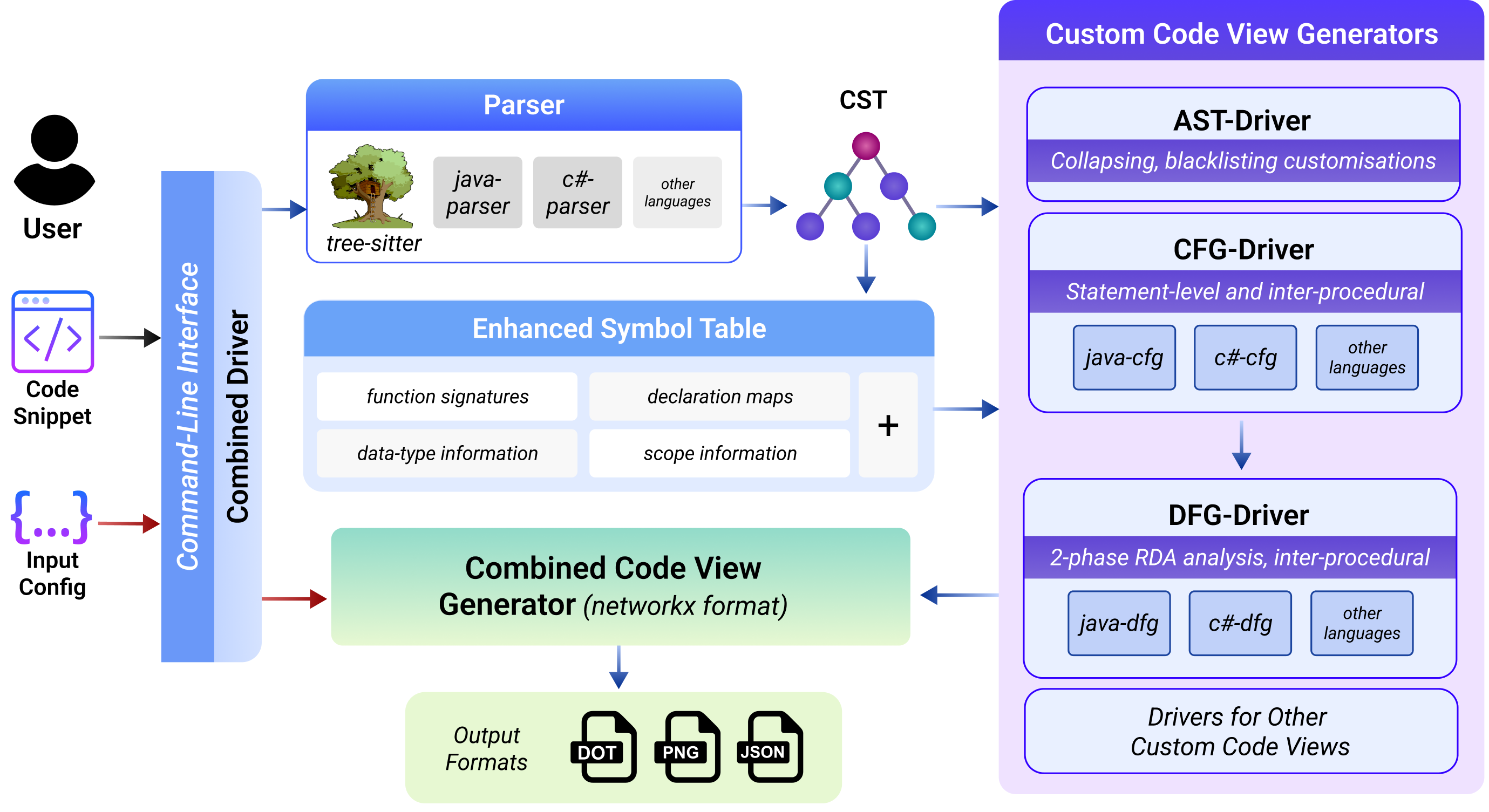}}
    \caption{Architecture of \texttt{COMEX}}
    \label{fig:arch_diag}
    \end{figure*}
Several ML4SE works leverage code-views such as the AST \cite{johnson2020learning}, the CFG \cite{bieber2022static}, the DFG  \cite{guo2020graphcodebert}, and their combinations (CDFG \cite{vasudevan2021learning}), to learn better code representations and improve performance on downstream SE tasks \cite{guo2020graphcodebert}.  
Unfortunately, most available tools that create such views are specific to a single language. 
\textit{SOOT} \cite{soot}, a popular static analysis tool for Java, requires the input Java code to be compilable and for all definitions to be available. But many existing research datasets are mostly method-level datasets with incomplete snippets and definitions \cite{codesearchnet, svajlenko2014towards}. Although \textit{python\_graphs} \cite{bieber2022library}, a framework for generating program graphs for Python, provides a composite ``program graph" with combined information from various typical code-views, it does not provide users the flexibility to combine, reduce or customize the typical code-views as supported by \texttt{COMEX}. \textit{Joern} is an open-source static analysis tool often used 
as a source for intermediate graph representations of code \cite{zhou2019devign, alsulami2017source, li2016vulpecker, dauber2018git, machiry2020spider} with support for Java, Python, C, C++, etc., providing code-views without a means to customize, combine, or easily extend to other languages. It has limited support for inter-procedural control-flow and data-flow analysis, and for interactive exploration and visualization\footnote{\href{https://galois.com/blog/2022/08/mate-interactive-program-analysis-with-code-property-graphs/}{https://galois.com/blog/2022/08/mate-interactive-program-analysis-with-code-property-graphs/}}. \texttt{COMEX} overcomes these limitations by providing support for generation of code-views through static code analysis even for non-compilable code both at function and program level, supporting out-of-the-box composition of views and easy extension to new languages without introducing further language-specific parser dependencies.

\section{The \texttt{COMEX} Package}

\begin{figure*}[h]
\centering
\hfill
{\includegraphics[width=\textwidth]{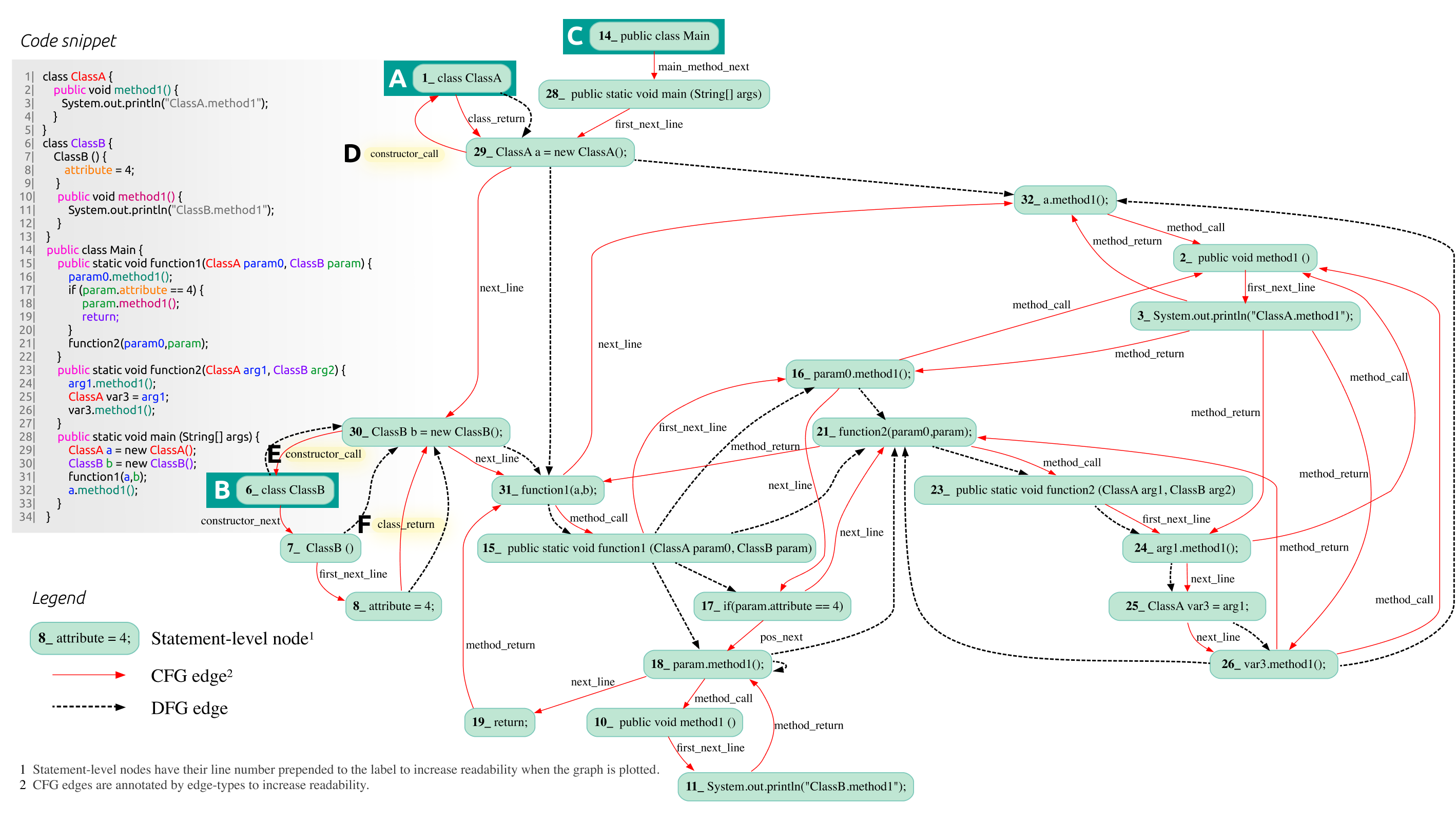}}
\caption{Statement-level CFG+DFG generated by \texttt{COMEX} for a code snippet with multiple functions showing inter-procedural control-flow and data-flow}
\label{fig:interprocedural}
\end{figure*}


\texttt{COMEX} is open-sourced\footnote{\url{https://github.com/IBM/tree-sitter-codeviews}} and also made available as a Python package\footnote{\url{https://pypi.org/project/comex/}}. Additionally, we have exposed a command-line-interface that allows users to conveniently specify the input code-snippet, output format types (dot,json,png) and any required customizations or combinations of different code-views. 
An overview of \texttt{COMEX} is depicted in Fig. \ref{fig:arch_diag}. As can be seen, \texttt{COMEX} starts with a code snippet and user-defined configuration as input. The snippet is then passed through a \textit{tree-sitter} parser to generate a concrete syntax tree (CST). An enhanced symbol table is created by processing the CST, and both of these together are used to create a CFG. Using the CFG, we implement reaching definition analysis (RDA) to generate the DFG. It is important to note that for CFG and DFG we implement both intra-procedural and inter-procedural analysis. In what follows, we elaborate on the details of the different code-views that we make available through \texttt{COMEX}.

\subsection{Abstract Syntax Tree}
We generate an AST by filtering some of the CST nodes provided by \textit{tree-sitter}. Trivial nodes such as semicolons (;) and braces (\{,\}) are dropped, while non-trivial nodes 
such as \textit{field\_access} or \textit{method\_invocation} are retained. 
We also provide customizations for the AST like (i) a \textit{collapsed AST} and (ii) a \textit{minimized AST}. A \emph{‘collapsed AST’} is one where all occurrences of the same variable are collapsed into one node. Whereas, in a \emph{‘minimized AST’}, certain node types can be \emph{‘blacklisted’} based on the purpose of the code representation. The rationale behind these customizations is to provide smaller ASTs without losing out on critical information. This results in fewer AST nodes, thus reducing graph sizes which helps make Graph Neural Network (GNN) \cite{long2022multi} approaches to source code representation learning computationally tractable. 

\subsection{Control-Flow Graph}
\label{cfg}

\textbf{\textit{Statement-level control-flow}} - Using the \textit{tree-sitter} generated CST and the enhanced symbol table, we proceed to create our CFG code-view. A typical CFG consists of a network of basic blocks, where each block is a set of instructions that execute sequentially with no intermediate control jump. Hence, constructing a CFG is usually a two-step process, where we first identify the basic blocks and then determine the control-flow edges between them.
However, in \texttt{COMEX}, we choose to produce a statement-level CFG that maps the control-flow between statements (and not blocks). This is useful for certain ML-based approaches and for generating the DFG as elaborated in (§\ref{dfg}).
The CFG for both Java and C\# is a statement-level approximation of control-flow.

\textbf{\textit{Inter-procedural control-flow}} - We support inter-procedural control-flow by statically analyzing all class definitions, object reference declarations, abstraction and inheritance specifications, method and constructor signatures and overloading. Fig. \ref{fig:interprocedural} shows a code snippet with two class definitions, \textit{ClassA} (\textbf{A}) and \textit{ClassB} (\textbf{B}), apart from the \textit{Main} class (\textbf{C}). The CFG edges are highlighted in red. The diagram depicts the change of control-flow during object instantiation to the corresponding class definition via ``constructor\_call" edges \textbf{D} ($29 \rightarrow 1$) and \textbf{E} ($30 \rightarrow 6$). As an explicit constructor is available for ClassB, the control flows through the constructor before returning to the site of instantiation via the ``class\_return" edge \textbf{F} ($8 \rightarrow 30$) . In case of method or constructor overloading, the function signatures are compared to determine the control-flow edges. When methods are called on object references, they are linked with the corresponding definition by matching the function signatures and available static references within the corresponding class. Nested function calls are also handled by tracking and mapping back all statically available signatures of function calls and their definitions.


\subsection{Data-Flow Graph}
\label{dfg}

Using the CFG generated in (§\ref{cfg}), we perform data-flow analysis to create our DFG code-view.
One of the fundamental techniques in data-flow analysis is \textit{Reaching Definition Analysis (RDA)}
where we identify the set of definitions that may reach a program point, i.e., the definitions that may affect the value of a variable at that point. A statement-level DFG is then generated using this information. Using the RDA-based implementation addresses many of the significant drawbacks that we found in the data-flow extraction logic used by GraphCodeBERT \cite{guo2020graphcodebert} such as lack of inter-procedural analysis, incorrect handling of scope information as well as data-flow thorough loops. It should be noted that the RDA-based analysis is inherently more computationally expensive.

In addition to method level analysis, we also support an out-of-the-box program-level DFG via a \textit{two-phase RDA}. The first phase is the typical RDA algorithm for each method, followed by another iteration of RDA that also takes into consideration the inter-procedural control-flow. This implementation helps track changes made to variables that are passed as parameters via method invocations. This is only performed for non-primitive data-types since primitive data-types are passed by value in Java and C\#. A full-blown alias analysis, which precisely determines all possible aliasing relationships can be challenging and computationally expensive. We hence support a partial alias analysis technique that approximates the possible memory references in a program. We also provide two additional data-flow relations - ``LastDef" and ``LastUse". Enabling ``LastDef" results in edges that link between re-definitions of variables as well as edges between declarations and definitions of variables. Similarly, ``LastUse" links the current use of a variable to the last program point where it was read. These relationships help add more edges in those method-level snippets that mainly use global variables which are not defined in the method body.


\subsection{Combinations and Customizations}

In addition to \textit{generating} code-views, \texttt{COMEX} can also \textit{combine and customize} multiple code-views into a single graph. 
For example, a combination of CFG and DFG would generate the two code-views separately and then combine them based on unique node identifiers as shown in Fig. \ref{fig:interprocedural}. Additionally, as we used just one parser package, we are able to implement this feature using a single module (\textit{CombinedDriver}) that works seamlessly across all languages. \texttt{COMEX} is currently capable of generating over $15$ different \textit{customized} representations\footnote{Please refer to \href{https://github.com/IBM/tree-sitter-codeviews/blob/main/List_Of_Views.pdf}{List-Of-Views.pdf} in the repository for a complete list}. 

\section{Discussion and Limitations}
\texttt{COMEX} was tested for robustness by generating and validating the code-views obtained on the large datasets popularly used for benchmarking ML-based SE tasks (CodeNet \cite{ibmcodenet}, CodeSearchNet \cite{codesearchnet} and \cite{translation}). Many of these datapoints have missing definitions and are not compilable, but their code-views were successfully generated as long as they were free of syntax errors. However, we are unable to provide a very accurate alias analysis that usually works only for compilable code because we support non-compilable input code snippets. Instead we provide a partial alias analysis. Among the aforementioned datasets, only \cite{translation} has C\# datapoints which is why we expect our implementation of Java code-views to be more robust than our C\# implementation. 

\section{Conclusion and Future work}
In source code representation learning research, there are many notable works that exploit code-specific properties like control-flow, data-flow, read-write dependencies, etc., in addition to treating code as regular natural language text. To this end, we believe that \texttt{COMEX} will enable researchers and developers in this domain to extract and customize structural information from code-views for new methods of representation learning. \texttt{COMEX} provides a framework which can be extended to support more code-views and their combinations and can be easily extended to many other popular languages like Python and C++ which can spur research in ML4SE and effective source code representation learning.







\bibliographystyle{IEEEtran}
\bibliography{IEEEabrv,references}



\end{document}